\documentclass[%
superscriptaddress,
reprint,
nofootinbib,
 amsmath,amssymb,
aps,
pra,
]{revtex4-1}

\usepackage[T1]{fontenc}
\usepackage{natbib}
\usepackage{graphicx}
\usepackage{dcolumn}
\usepackage{bm}
\usepackage{verbatim} 
\usepackage[
    colorlinks=true,
    linkcolor=blue,
    urlcolor=blue]{hyperref}
\usepackage{xspace} 
\usepackage[capitalize]{cleveref}
\crefname{section}{Sec.}{Secs.}
\Crefname{section}{Sec.}{Secs.}
\crefname{appendix}{App.}{Apps.}
\Crefname{appendix}{App.}{Apps.}
\Crefname{figure}{Fig.}{Figs.}
\crefname{figure}{Fig.}{Figs.}
\creflabelformat{equation}{(#2#1#3)}

\usepackage{orcidlink}

\usepackage{xcolor}

\newcommand\blfootnote[1]{%
  \begingroup

\renewcommand\thefootnote{}\footnote{\hspace{-1.5em}#1}%
  \addtocounter{footnote}{-1}%
  \endgroup
}

\begin{document}

\title{Floquet engineering of spin-spin interactions in a hybrid atomic system}

\author{Daniel Gavilan-Martin\orcidlink{0009-0007-9321-1876}}
\email{gaviland@uni-mainz.de}

\affiliation{Johannes Gutenberg-Universit\"at Mainz, 55128 Mainz, Germany}
\affiliation{Helmholtz Institute Mainz, 55099 Mainz, Germany} 
\affiliation{GSI Helmholtzzentrum für Schwerionenforschung GmbH, 64291 Darmstadt, Germany}

\author{Grzegorz Łukasiewicz\orcidlink{0000-0001-5820-1671}}
\email{grzegorz.lukasiewicz@doctoral.uj.edu.pl}

\affiliation{Marian Smoluchowski Institute of Physics, Jagiellonian University in Krak\'ow, Łojasiewicza 11, 30-348, Krak\'ow, Poland}
\affiliation{Doctoral School of Exact and Natural Sciences, Jagiellonian University in Krak\'ow, Łojasiewicza 11, 30-348, Krak\'ow, Poland}

\author{Vincent Kai Sch\"afer\orcidlink{0009-0000-2840-0450}}
\affiliation{Johannes Gutenberg-Universit\"at Mainz, 55128 Mainz, Germany}
\affiliation{Helmholtz Institute Mainz, 55099 Mainz, Germany}

\author{Mikhail Padniuk\orcidlink{0000-0002-8709-143X}}
\affiliation{Arda Atomics GmbH, Epplestraße 225, 70567 Stuttgart, Germany}

\author{Adam Stefa\'nski\orcidlink{0009-0007-4910-6307}}
\affiliation{Marian Smoluchowski Institute of Physics, Jagiellonian University in Krak\'ow, Łojasiewicza 11, 30-348, Krak\'ow, Poland}

\author{Adam W\k{e}glik\orcidlink{0009-0005-4172-2995}}
\affiliation{Marian Smoluchowski Institute of Physics, Jagiellonian University in Krak\'ow, Łojasiewicza 11, 30-348, Krak\'ow, Poland}

\author{Emmanuel Klinger\orcidlink{0000-0001-7461-9568}}
 \affiliation{Universit\'e Marie et Louis Pasteur, Supmicrotech, CNRS, Institut FEMTO-ST, 25000 Besan\c{c}on, France}

\author{Szymon Pustelny\orcidlink{0000-0003-3764-1234}}
\affiliation{Marian Smoluchowski Institute of Physics, Jagiellonian University in Krak\'ow, Łojasiewicza 11, 30-348, Krak\'ow, Poland}

\author{Derek F. Jackson Kimball\orcidlink{0000-0003-2479-6034}}
\affiliation{Department of Physics, California State University - East Bay, Hayward, CA 94542, USA}
 
\author{Dmitry Budker\orcidlink{0000-0002-7356-4814}}
\affiliation{Johannes Gutenberg-Universit\"at Mainz, 55128 Mainz, Germany}
\affiliation{Helmholtz Institute Mainz, 55099 Mainz, Germany} 
\affiliation{GSI Helmholtzzentrum für Schwerionenforschung GmbH, 64291 Darmstadt, Germany}
\affiliation{Department of Physics, University of California, Berkeley, CA 94720, USA}
 
\author{Arne Wickenbrock\orcidlink{0000-0001-5540-7519}}
\affiliation{Johannes Gutenberg-Universit\"at Mainz, 55128 Mainz, Germany}
\affiliation{Helmholtz Institute Mainz, 55099 Mainz, Germany} 
\affiliation{GSI Helmholtzzentrum für Schwerionenforschung GmbH, 64291 Darmstadt, Germany}

\blfootnote{$^{*,\dagger}$\,These authors contributed equally to this work.}

\begin{abstract}
We demonstrate dynamical control of the effective spin-spin interaction, dominated by Fermi-contact interaction, in a hybrid spin system via parametric modulation. We show that, in an alkali-noble-gas comagnetometer, periodic modulation of the direction of the electron spin polarization with respect to the nuclear polarization leads to a Floquet-induced renormalization of the spin-exchange coupling, governed by a zeroth-order Bessel function. This effect enables continuous tuning and suppression of the effective interaction strength without altering the intrinsic properties of the system. We develop a theoretical model that supports the experimental measurements. The results establish a general mechanism for controlling interaction strengths in hybrid atomic systems and provide new opportunities for precision measurements and quantum memories.
\end{abstract}

\maketitle

\section{Introduction}
Noble-gas nuclear spins are among the most robust quantum systems. They are shielded by fully filled electronic orbitals and are weakly coupled to the environment, leading to exceptionally long coherence times, often reaching many hours or even days \cite{HEIL1995337}. Noble gas spin ensembles can be confined in room-temperature vapor cells without the need for cryogenic infrastructure, making them an attractive platform for precision measurements \cite{safronova_search_2018,jackson2023probing,cong2025spin} and development of quantum technologies \cite{katz2022optical,katz2025optimal}. For example, noble gases are used for fundamental physics, including tests of Lorentz invariance \cite{vasilakis_limits_2009,Allmendinger:2013eya} and searches for new particles \cite{kornack_thomas_test_2005,lee_laboratory_2022,gavilan-martin_searching_2025,bloch_nasduck_2022,bloch_axion-like_2020,jiang_search_2021,zhang2023search}, metrology and rotation sensing \cite{kornack_nuclear_2005, wei_ultrasensitive_2023}. Noble gases have also been proposed to realize optically-addressed quantum memories \cite{katz_coupling_2021,katz_quantum_2022,Barbosa:2024vlt} and spin squeezing \cite{serafin2021squeezing}.

Effective ways of initializing, manipulating and reading-out the nuclear spin state are required for these applications. However, the weak coupling that allows for long-lived coherence is at the same time an obstacle to accessing the nuclear spins.  
Optical transitions can be used to manipulate the nuclear spin via the electronic structure. For noble gases, these are often in the UV regime, where finding an appropriate light source may be challenging. Another solution is to include an alkali metal in the vapor cell.
Noble-gas atoms collide with alkali-metal atoms, and they can exchange angular momentum \cite{walker1997spin,walker2011fundamentals,appelt_theory_1998}. The transfer of spin orientation in collisions is mediated by a dipole-dipole interaction dominated by Fermi-contact interaction \cite{Walker1989PhysRevA}. Typically, the spin-exchange collision rate is much faster than the other dynamics of the system \cite{happer_effect_1977,savukov_effects_2005}. As a result, individual collisions cannot be resolved.
Instead, it is the average behavior in time and over the volume of the vapor cell containing the mixture that can be manipulated and read out. The strength of the effective interaction depends, among other things, on the angle between the average magnetic moments of the different species. This means that controlling the relative spin orientation of the two atomic ensembles provides a direct way to manipulate their effective interaction.

The interaction rate can be manipulated in several ways. 
The vapor parameters, such as pressure and temperature, can be tuned to increase the collision rate. 
Overall faster spin transfer can also be achieved by increasing the differential polarization between the alkali atoms and noble gas atoms.
Additionally, tuning a DC magnetic field to match the precession frequencies of the alkali metal and noble gas increases the spin-transfer rate. When this condition is fulfilled a hybrid resonance is reached, and both ensembles have to be treated in a coupled basis \cite{shaham2022strong}. This is at the core of the operational principle of many noble-gas-based sensors \cite{su2022review}. 

\begin{figure*}[htb]
    \centering
    \includegraphics[width=\linewidth]{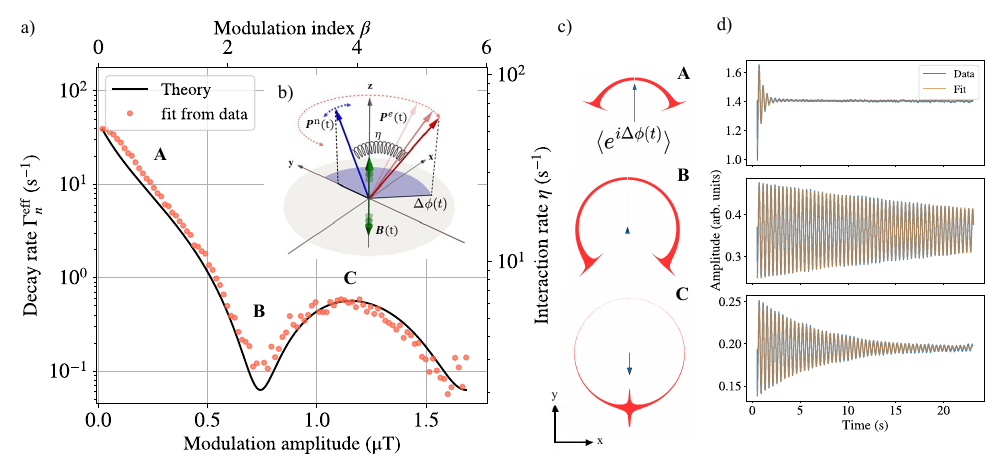}
    \caption{Floquet engineering of the spin-spin interaction. 
    The inset in the left plot [b)] shows a sketch of the effect of the parametric modulation of the leading field in the interaction between electronic and nuclear spins. The modulated magnetic field (in green) causes both nuclear ($P^n$, in blue) and electronic ($P^e$, in red) polarization to precess in the transversal plane. Since the gyromagnetic ratio of the nuclear spins is much smaller than that of the electron spins, $\gamma_n\ll \gamma_e$, the effect of the modulation for the nuclear spins is almost negligible, and the relative angle between polarizations can be changed. This affects the symmetrized coupling between them ($\eta$, represented with a spring), which can be tuned via the amplitude and frequency of the modulation. Subset c) shows an example of the orientation probability of the electronic polarization over one cycle of the modulation, for different values of the modulation amplitude. Since the modulation is much faster than the nuclear and electronic Larmor frequencies, the relative angle is averaged, leading to a Bessel function behavior of the interaction $\eta$, see \cref{eq:effective_matrix}. Free-precession-decays (FPD) corresponding to the respective electronic orientation probabilities are shown in d), where we can observe that the relaxation rate of the FPD, $\Gamma_n^\mathrm{eff}$, proportional to $\eta$, changes for different values of the modulation. The relaxation rate $\Gamma_n^\mathrm{eff}$ depends on $J_0^2(\beta)$, and follows the theory projection (black solid line) in a), according to \cref{sec:theory}. 
    }
    \label{fig:Interaction rate cartoon}
\end{figure*}

Parametric modulation with an AC magnetic field provides further control over the spin dynamics in such systems \cite{haroche1970modified}. For example, magnetic-field modulation has been used in alkali-metal magnetometers for lock-in detection, shifting signals to higher frequencies, which reduces the impact of low-frequency technical noise \cite{li_parametric_2006, parametrically_2018_Jiang,put2019nonlinear}. More recently, parametric modulation techniques have been explored as tools for engineering spin dynamics and modifying the system response through Floquet-like control \cite{jiang_floquet_2021, jiang_floquet_2022,bloch_nasduck_2022,GOLUB19941,Tat:2025dyl}. These approaches have enabled dual-axis sensitivity and new signal-processing schemes in atomic-spin-based sensors.
 
In this work, we apply Floquet engineering techniques to manipulate the spin-spin interaction with a periodic driving. In particular, we investigate the modulation of the leading magnetic field in an alkali-metal-noble-gas mixture. We demonstrate that such modulation can be used to dynamically control the effective interaction between the electronic (alkali metal) and nuclear (noble gas) spin ensembles. 

By changing the modulation parameters, one can dynamically alternate between a strongly and a weakly coupled regime. This provides a new tool for manipulation of coupled spin systems. Specifically, interaction rates can be tuned independently of the Larmor frequency of the spin ensembles, allowing a new degree of freedom that is not present with DC-field manipulation, see \cref{fig:Interaction rate cartoon}. 

This paper is structured as follows. The theory of the effect is presented in \cref{sec:theory}, followed by a presentation of experimental results in Sec.\,\ref{sec:results}. Finally, the implications of the results and future directions are discussed in Sec.\,\ref{sec:discussion}. 

\section{Theory} \label{sec:theory}
In this section, we consider an overlapping ensemble of alkali-metal atoms with electronic polarization and a noble gas with nuclear polarization. They are confined in a gas cell and can be manipulated with laser light or magnetic fields. The interaction between both ensembles is given by the microscopic dipole-dipole interaction Hamiltonian between the nuclear spin $\mathbf{I}$ of the noble gas and the electron spin $\mathbf{S}$ of the alkali metal. The interaction is dominated by the Fermi-contact interaction, which can be written as $\mathcal{H}_{\textrm{FC}} = \lambda\,\mathbf{S}\cdot \mathbf{I}$, where $\lambda$ is the coupling parameter \cite{happer_spin-exchange_1973, appelt_theory_1998,Walker1989PhysRevA}. Here, we neglect the hyperfine structure of the alkali metal and focus on its interaction with the noble-gas nuclear spin. In the mean-field approximation, this interaction is equivalently described as the mutual effective magnetic fields produced by one ensemble acting on the other, where the Fermi-contact interaction enhances and adds to the underlying magnetic coupling.

We define the $z$ axis as the direction of the optical pumping and applied leading magnetic field $B_z$, and refer to it as the longitudinal direction.
In the following, the indices $e$ and $n$ are used consistently to label quantities referring to the alkali-metal electron ensemble and the noble-gas nuclear ensemble, respectively.
We work in the small-angle approximation, assuming that the polarization vectors $\mathbf{P^e}$ and $\mathbf{P^n}$ remain close to the longitudinal direction.  
In this picture, the macroscopic interaction between $\mathbf{P^e}$ and  $\mathbf{P^n}$ can be classified into two categories:
\begin{enumerate}
    \item Longitudinal interaction, which produces static shifts of the electron and nuclear Larmor frequencies through the effective fields $B^n_\parallel~=~\lambda_\parallel M^n P^n_\parallel$ and $B^e_\parallel = \lambda_\parallel M^e P^e_\parallel$ corresponding to the longitudinal magnetizations $M^n P^n_\parallel$ and $M^e P^e_\parallel$ enhanced by Fermi-contact interaction factor $\lambda_\parallel$.
    
    \item Transverse exchange interaction, which dynamically couples the transverse polarization components of the two ensembles with coupling factors 
    \begin{equation}
    \label{eq:interaction}
        \eta_{en} = \frac{\gamma_e}{q} \lambda_\perp M^n P^e_{\parallel}\,, \ \ 
        \eta_{ne} = \gamma_n \lambda_\perp M^e P^n_{\parallel}\,.
    \end{equation}
\end{enumerate}
Here $\gamma_e$ and $\gamma_n$ denote respective gyromagnetic ratios and $q$ is the slowing-down factor that accounts for the hyperfine coupling between the electron and nuclear spins of the alkali atom,  which reduces the effective precession rate of the electron spin in the spin-exchange-relaxation-free (SERF) regime \cite{allred2002high}.
We distinguish the collisional enhancement factor $\lambda_\parallel$, entering longitudinal polarization components, from the coupling governing the transverse exchange $\lambda_\perp$. Under ordinary conditions the interaction is isotropic, and the two couplings are equal, such that $\lambda_\perp = \lambda_\parallel = \lambda$ \cite{walker1997spin}. The distinction between these two factors will prove useful when discussing the effect of parametric modulation on the transverse spin dynamics. 

Taking both effects into account, the transverse polarization dynamics are described by the coupled Bloch-Hasegawa equations. We follow the formalism of Ref.~\cite{kornack_dynamics_2002,shaham_strong_2022,happer2010optically}:
\begin{equation}
\label{eq:Bloch_hasegava_compact}
\begin{pmatrix}
\dot{P}^e_{\perp} \\
\dot{P}^n_{\perp}
\end{pmatrix}
=
-iM(t)
\begin{pmatrix}
P^e_{\perp} \\
P^n_{\perp}
\end{pmatrix}
-i
F(t)\,,
\end{equation}
where $P_\perp=P_x+iP_y$, $M(t)$ is the interaction matrix. and $
F(t) =  
[\frac{\gamma_e}{q} \alpha_e b_{\perp}(t) P_{\parallel e}\,,\ \gamma_n \alpha_n b_{\perp}(t) P_{\parallel n}]^T
$
describes the external perturbation, due to a magnetic or pseudomagnetic interaction of amplitude $b_\perp (t)$, coupling to electronic and nuclear polarizations with strengths $\alpha_e$ and $\alpha_n$, respectively (we follow the convention introduced in Ref. \cite{padniuk_universal_2024}).

The real part of the diagonal elements of the interaction matrix consist of the mean Larmor frequencies  due to the \textit{static} magnetic fields $\omega_e = \frac{\gamma_e}{q}(B_z + B^n_\parallel)$ and $\omega_n~=~\gamma_n(B_z + B^e_\parallel)$, as well as 
an \textit{oscillating} component due to the modulation of the leading field, $B_\textrm{m}\cos(\omega_\textrm{m}t)$, which we write explicitly. The imaginary part of the diagonal elements are the respective relaxation rates $\Gamma_e$ and $\Gamma_n$. By defining $\Omega_e = \frac{\gamma_e}{q}B_\text{m}$ and $\Omega_n = \gamma_n B_\text{m}$, the interaction matrix is: 
\begin{equation}
\begin{split}
\label{eq:interaction_matrix}
&M(t) =\\
&\begin{pmatrix}
\omega_e + \Omega_e \cos(\omega_\textrm{m}t) - i \Gamma_e &  \eta_{en} \\
\eta_{ne} & \omega_n + \Omega_n\cos(\omega_\textrm{m} t) - i\Gamma_n
\end{pmatrix}\,.
\end{split}
\end{equation}
The off-diagonal terms represent the dynamical coupling between the transverse polarizations of the electron and nuclear ensembles. 

The equations are simplified by going to the frame where the terms $\Omega_{e,n} \cos(\omega_m t)$ are canceled. In this frame, we average over the fast modulation cycle, retaining only dynamics slower than the modulation frequency. Details on the corresponding unitary transformation and the following derivation steps, as well as the  approximations performed are provided in \cref{app:matrix_derivation}.
Finally, we obtain the following effective interaction matrix:
\begin{equation}
\label{eq:effective_matrix}
    M_\textrm{eff}(\beta) =
\begin{pmatrix}
\omega_e - i\,\Gamma_e 
& \eta_{en}\, J_0(\beta) \\[6pt]
\eta_{ne}\, J_0(\beta) 
& \omega_n - i\,\Gamma_n
\end{pmatrix}\,,
\end{equation}
where $\beta = (\Omega_e - \Omega_n)/\omega_{\textrm{m}}\approx \Omega_e / \omega_\textrm{m}$ is the modulation index. Importantly, the modulation-induced factor $J_0(\beta)$ renormalizes only the coupling between the transverse polarization components defined in \cref{eq:interaction}, while the longitudinal interaction determining the static frequency shifts remains unaffected. It is equivalent to introducing an effective anisotropy of the collisional interaction, such that $\lambda^{\textrm{eff}}_\perp = J_0(\beta) \lambda_\perp$ and $\lambda^{\textrm{eff}}_\parallel = \lambda_\parallel$. 

The dynamics of the coupled system are naturally described in terms of the eigenmodes of the effective interaction matrix $M_\mathrm{eff}(\beta)$. The eigenvalues of $M_\mathrm{eff}(\beta)$ represent the oscillation frequencies and decay rates of the coupled spin modes. The read-out of the system projects the dynamics onto the electronic spin component (either $P_x^e$ or $P_y^e$), leading to a superposition of both eigenvalues. The complex eigenvalues are
\begin{equation}
\begin{split}
\label{eq:eigenvalues}
\xi_{\pm}(\beta)
=&\, \frac{\omega_e+\omega_n}{2}
 - i\frac{\Gamma_e + \Gamma_n}{2}
 \\ \pm &\,
 \sqrt{
 \frac{\left(\Delta \omega - i\Delta \Gamma\right)^2}{4}
 + \eta^2\,J^2_0(\beta)
 }\,,
\end{split}
\end{equation}
where $\Delta\omega = \omega_e - \omega_n$,  $\Delta\Gamma = \Gamma_e - \Gamma_n$, and $\eta^2 = \eta_{ne}\eta_{en}$.

The exact eigenvalues $\xi_{\pm}(\beta)$ fully describe the coupled dynamics of the system. 
In the regime of high modulation index, where 
\begin{equation}
\label{eq:high_mod_regime}
    \eta^2 J_0^2(\beta)\ll(\Delta\omega)^2+(\Delta\Gamma)^2\,,
\end{equation}
the coupling between the two spin ensembles is strongly suppressed, and the system is weakly hybridized. The eigenvalues can therefore be treated as perturbations of the bare nuclear-like ($n$) and electron-like ($e$) modes. Expanding \cref{eq:eigenvalues} to leading order in the small parameter $\eta^2 J_0^2(\beta)$, we obtain
\begin{equation}
\xi_{n(e)} \;\approx\;
\omega_{n(e)} - i\Gamma_{n(e)}
\;\mp\;
\frac{\eta^2 J_0^2(\beta)}{\Delta\omega - i\,\Delta\Gamma}\,.
\end{equation}
The electron-like mode is short-lived with a relaxation rate on the order of $\Gamma_e$, while the nuclear mode is long-lived ($\Gamma_n \ll \Gamma_e$). The nuclear effective relaxation is given by
\begin{equation}
    \Gamma_{n}^\textrm{eff} \approx -\textrm{Im}(\xi_n) = \Gamma_n + \frac{\eta^2 J_0^2(\beta) \Delta\Gamma}{(\Delta\omega)^2 + (\Delta \Gamma)^2}\,,
    \label{eq:lambda effective}
\end{equation}
where the second term is dominant. At the same time, the nuclear-like mode frequency is
\begin{equation}
    \omega_{n}^\textrm{eff} \approx \textrm{Re}(\xi_n) = \omega_n - \frac{\eta^2 J_0^2(\beta) \Delta \omega }{(\Delta\omega)^2 + (\Delta \Gamma)^2}\,.
    \label{eq:omega effective}
\end{equation}
Parametric modulation therefore provides a simple and continuous control over the effective coupling between the spin ensembles. By tuning the factor $J_0(\beta)$, the alkali-induced nuclear relaxation can be suppressed without altering the static detuning $\Delta \omega$. In particular, near the zeros of $J_0(\beta)$, the transverse coupling is effectively switched off, and the nuclear mode approaches its intrinsic lifetime and bare precession frequency, despite the presence of the polarized alkali ensemble.

Additional effects of parametric modulation on the alkali metal polarization and the signal demodulation at the first and second harmonic of modulation frequency are discussed in \cref{app:alkali metal modAmp}. Implications of the AC field on a self-compensating comagnetometer response function are provided in \cref{app:self-compensatig response}.

\section{Results} \label{sec:results}
We now experimentally verify the validity of the model presented in \cref{sec:theory}. Using a similar comagnetometer setup as described in Refs.\;\cite{klingerPhysRevApplied.19.044092,padniuk_universal_2024} (see Appendix\;\ref{app:ExpSetup} for experimental details), we have measured free precession decays (FPD) of the coupled system caused by a step change of the applied magnetic field in the transverse plane. The read-out is perfomed via Faraday rotation of linearly polarized light, directly proportional to the alkali-metal polarization and indirectly reflects the noble-gas dynamics through their coupling ($\eta$). We observe the reponse under different parameters of the leading field modulation. The output signal is demodulated at the first harmonic of the modulation of the leading field $\omega_\text{m}$, corresponding to a measurement of $P^e_x$, see \cref{app:alkali metal modAmp}. The measurements are performed with a K-Rb-$^3$He comagnetometer, where Rb is used to initialize the spin polarization, and the electron spin dynamics are governed by the much more abundant K, see \cref{app:ExpSetup}.

We  extract the decay rate of the nuclear-like FPD, which corresponds to the effective relaxation time \hbox{$T_2^{\mathrm{eff}} = 1/\Gamma_n^\mathrm{eff}$}. Under the experimental conditions listed in \cref{app:ExpSetup}, the inequality from \cref{eq:high_mod_regime} is fulfilled, so the effective relaxation time can be approximated by \cref{eq:lambda effective}. In the regime where $\Gamma_e \gg \Gamma_n$, the nuclear relaxation is dominated by the coupling to alkali spins, and $\Delta\Gamma \approx \Gamma_e$. The measured decay rate provides a direct measure of the symmetrized coupling strength $\eta J_0^2(\beta)$. The precession frequency and signal amplitude can be extracted similarly from an FPD. Its dependence on the parametric modulation parameters is explicit in \cref{eq:omega effective,eq:amplitude first harmonic}.

Given the rapid decay of the electron-like mode, we neglect its contribution and model the FPD as an exponentially damped oscillation,
\begin{equation}
f(t;A,\Gamma^\mathrm{eff}_n,\omega^\mathrm{eff}_n,\phi,b)=Ae^{-\Gamma^\mathrm{eff}_n t}\sin(\omega_n^\mathrm{eff} t+\phi)+b\,,
\label{eq:timeseries fit}
\end{equation}
where $A$ is the measured amplitude, $\Gamma^\mathrm{eff}_n$ and $\omega_n^\mathrm{eff}$ are the effective decay rate and frequency of nuclear spins, respectively, and $\phi$ and $b$ are angular and amplitude offsets. 

Figure \ref{fig:Interaction rate cartoon}d shows examples of FPD measurements with their corresponding fits for various values of the parametric modulation amplitude.  There is a non-monotonic dependence of the decay rate on the amplitude modulation, as we cross through a minimum $\Gamma^\mathrm{eff}_n$ around $0.75\,\mathrm{\mu T}$, as shown in \cref{fig:Interaction rate cartoon}a. This dependence, as shown in \cref{eq:lambda effective}, is the square of a zeroth-order Bessel function. 

We study the dependence on $B_m$ with fixed modulation index $\beta$ in \cref{app:high Bz}. At large $B_m$, both amplitude of the output signal and decay rate are suppressed approximately by $1/B_m^2$. After accounting for this effect, we compare the behavior of the system as a function of $\beta$ with the theoretical model. Note that only $\omega_m$ and $B_m$ can be directly experimentally manipulated. Since the modulation index $\beta \propto B_m$, we scan $B_m$ and extract $A$, $\Gamma^\mathrm{eff}_n$, and $\omega^\mathrm{eff}_n$ from the fits of the FPD with \cref{eq:timeseries fit}.
In particular, we are interested in the zeros of the Bessel functions, when the decoupling between nuclear and electron spins is maximized, see \cref{eq:lambda effective}. The amplitude and decay rate of the measured FPD are fit with the following expressions
\begin{subequations}
\label{eq:beta fits}
\begin{align}
    A&=C_aA(B_m)J_0(\beta)J_1(\beta)+b_A \,, \label{eq:modAmp}\\
    \Gamma_n^\mathrm{eff}&=C_\Gamma \Gamma^\mathrm{eff}_n (B_m)J_0(\beta)^2+b_\Gamma\,, \\
    \omega&=\omega_n-C_\omega J_0(\beta)^2 \,,
\end{align}
\end{subequations}
where $C_{i}$, with $i=\{a,\Gamma,\omega\}$, is a parameter that defines the maximum modulated amplitude, decay rate, and frequency, and $b_i$ is an offset.

The zeroes of the decay rate and amplitude are predicted from the applied modulation index $\beta$, up to the exact value of the slowing-down factor $q$. By matching the measured zeroes with the expected zeroes of the Bessel functions of both amplitude and decay rate simultaneously we find that the best fit for the effective electron gyromagnetic ratio is $\gamma_e/q= 2\pi \times 6.7$\,Hz/nT, with $q=4.17$ when applying $\omega_m/2\pi=2\,\mathrm{kHz}$.
For $^{39}$K vapor, the slowing-down factor $q$ ranges from 4-6, depending on the polarization of the sample and the alkali-alkali spin-exchange rate \cite{savukov_effects_2005}.
Then, $C_i$ and $b_i$ are fit.

\begin{figure}[h]
    \centering
    \includegraphics[width=\linewidth]{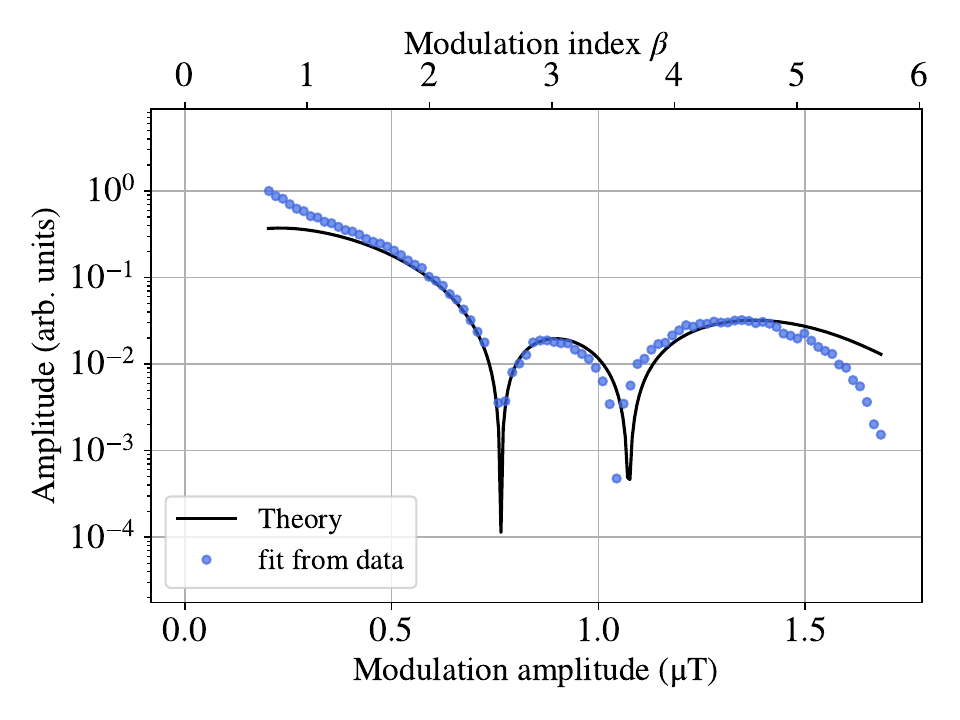}
    \caption{Amplitude response of the system (x-axis). The transverse polarization is modulated, giving accessed to both directions transverse to the leading field (z). We take into account the reduction of the amplitude due to the high AC magnetic field applied, see \cref{app:high Bz}, $A(B_m)$. The expression fitted for the amplitude is \cref{eq:modAmp} , and is further discussed in \cref{app:alkali metal modAmp}. Due to the dependence in $J_1(\beta)$ there is an additional zero not present in \cref{fig:Interaction rate cartoon} (right).}
    \label{fig:AmpModulated}
\end{figure}

In \cref{fig:Interaction rate cartoon}a, the effective nuclear decay rate $\Gamma_n^{\mathrm{eff}}$ 
is shown as a function of the modulation amplitude. The equivalent values of the modulation index $\beta$ are also displayed.
Up to almost $2\,\mathrm{\mu T}$, two zeros are clearly visible, which also match the amplitude in \cref{fig:AmpModulated}. The latter figure presents an additional zero due to its dependence on $J_1(\beta)$ [see \cref{app:alkali metal modAmp}].

An additional FPD measurement is performed to explore higher $\beta$ for the measured nuclear spin frequency $\omega^\mathrm{eff}_n$. In \cref{fig:omegaNeff}, $\omega^\mathrm{eff}_n$ never reaches zero, but rather fluctuates periodically around the Larmor frequency $\omega_n$ as a function of a zeroth-order Bessel function. 
The effective electron gyromagnetic ratio $\gamma_e/q=5.7(1)\,$Hz/nT is independently measured. Up to $\beta=11$ is reached, and the model reproduces the data.

In all three measured quantities, namely $\Gamma_n^{\mathrm{eff}}$, $\omega_n^{\mathrm{eff}}$ and amplitude of the response, there is a relatively small systematic mismatch between the 
measured parameters and the theory predictions. These systematic deviations could be attributed to two main effects. At low modulation indices, the contribution of the rapidly decaying electron-like mode to the FPD is not fully negligible. At larger modulation amplitudes, changes in relaxation modify the effective slowing-down factor $q$ \cite{savukov_effects_2005}, distorting the relation $\beta \approx (\gamma_e/q)B_\text{m}/\omega_\text{m}$. While these effects lie beyond the scope of the present model, they do not pose a limitation for practical applications once the coupled and decoupled regimes are experimentally calibrated.

The interaction strength $\eta$ is inferred from the measured decay rate, according to \cref{eq:lambda effective}. In particular, we measure $\Gamma_e= 10\,\mathrm{s}^{-1}$ and $\Delta \omega/2\pi=- 2.2\,\mathrm{Hz}$. We demonstrate control of the interaction parameter $\eta$ over more than an order of magnitude, in good agreement with the expected zeroth-order Bessel-function dependence. This corresponds to a variation of the nuclear decay rate over nearly two orders of magnitude, with a minimum coupling at $\beta_{\mathrm{min}} \approx 2.4$.

\begin{figure}[h]
    \centering
    \includegraphics[width=\linewidth]{Figures/omegaNeff.pdf}
    \caption{
    Effective frequency of the nuclear spin precession $\omega_n/2\pi$. In green, the experimental values extracted from FPD fits are displayed. The frequency rises from below 2.80\,Hz up to 2.95\,Hz, and oscillates from $\beta\geq 2$ in function of the modulation index $\beta$. The theory prediction is displayed in black. The slowing down factor is independently measured such that $\gamma_e/q=5.7(1)\,$Hz/nT.
    }
    \label{fig:omegaNeff}
\end{figure}

\section{Discussion} \label{sec:discussion}
In this work, we demonstrate that the modulation of the leading magnetic field provides a direct method for controlling the 
spin interaction in an alkali-metal-noble-gas spin ensemble. On average, the AC field modifies the spin-exchange coupling in the perpendicular plane to the modulation by a zeroth-order Bessel function $J_0(\beta)$, while leaving the longitudinal interaction unchanged. The transverse coupling between electronic and nuclear spin ensembles can be tuned by adjusting the modulation index $\beta$, which depends on the amplitude and frequency of the applied modulation field.

This effect is observed in the measured free-decay signals. The nuclear relaxation rate and frequency shift follow the predicted $J_0^2(\beta)$ dependence as the modulation index $\beta$ varies. The interaction between the two ensembles is strongly suppressed near the zeros of the Bessel function. In this regime, the transverse coupling is effectively removed, leading to almost-free nuclear spin dynamics while maintaining the same static magnetic configuration. 
The nuclear spins can remain coherent for hours, and their quantum state is effectively stored.

Strong interactions can be used to initialize and read out the nuclear spins, recovering the information, as already proposed with static fields \cite{katz_coupling_2021,shaham_strong_2022}. Floquet-engineering becomes a tool to dynamically alternate into the different states of a quantum memory, or, if the nuclear spins can be modified during the storage time, a quantum processor.
  
Furthermore, the effective two-level interaction matrix in \cref{eq:effective_matrix}, which describes the transverse polarization dynamics, provides a natural platform for exploring non-Hermitian degeneracies known as exceptional points. In this system, an exceptional point occurs when the two complex eigenvalues of the matrix [see \cref{eq:eigenvalues}] become equal while their corresponding eigenvectors simultaneously coalesce, leading to a non-diagonalizable matrix and a complete merging of the dynamical response. Exceptional points have been extensively studied in other systems~\cite{miri_exceptional_2019, bergholtz_exceptional_2021}, and recently in the context of hot atomic vapors \cite{liang_observation_2023,ellis_systematic_2025, wei_phase_2025, kopciuch_liouvillian_2025}. This direction of research has attracted significant interest due to their potential applications in precision measurements. Near the exceptional point, the response of the system is nonlinear in the control parameter, and the signal can be enhanced. However, it is currently under active debate if this leads to increased SNR in a realistic setting~\cite{wiersig_review_2020,Wang:2025lcm}. In this context, parametric modulation can be used to tune the system across the exceptional point, and may be particularly useful for implementing protocols such as exceptional-point encircling~\cite{dembowski_encircling_2004}.

As shown in \cref{sec:results}, for a modulation frequency $\omega_m/2\pi= 2$\,kHz, a high modulation amplitude compared to $B_c$ needs to be applied to reach the decoupled regime, i.e. the first zero of the Bessel function. This leads to suppression in the amplitude response of the system due to the broadening of the K magnetic resonance, see \cref{app:high Bz}.
This could be avoided by lowering the modulation frequency, requiring lower modulation amplitude for the same $\beta$,
 effectively increasing signal-to-noise-ratio (SNR).

While so far we have focused on reaching the hybrid resonance and then suppressing the interaction as a way to demonstrate its tunability, the complementary proposition is intriguing: use the modulation not to suppress the interaction but to enhance it. Static fields can be set up to maximize the stability of the system, even far detuned from the hybrid resonance. In this context, the assumptions made in \cref{sec:theory}, where a regime close to the hybrid resonance is assumed, are no longer valid. We generalize the mathematical framework for the far detuned regime in \cref{app:off-resonance-coupling}, and calculate the eigenvalues of the system beyond the slow dynamic approximation. This theoretical description is in direct connection with recent modulation-based comagnetometers and spin masers \cite{jiang_floquet_2022, bloch_nasduck_2022}, where longitudinal driving generates Bessel-weighted sidebands and enables resonant processes away from the bare frequencies.

While the emphases have usually been on the sensing possibilities, information can be coherently exchanged and stored in nuclear spin states. Floquet engineering of the interaction might make hybrid atomic systems a versatile tool in quantum information processing. 

\section*{Acknowledgements}
The authors thank Martin Jutisz, Marek Kopciuch, Jason Stalnaker, Ibrahim Sulai, Giorgos Vasilakis, and Weiyi Wang for useful discussions. 
This work has been supported by the Cluster of Excellence ``Precision Physics, Fundamental Interactions, and Structure of Matter'' (PRISMA++ EXC 2118/2) funded by the
German Research Foundation (DFG) within the German Excellence Strategy (Project ID 390831469). SP acknowledges the support from the National Science Center, Poland within the OPUS program (2020/39/B/ST2/01524). GL acknowledges support from the National Science Center, Poland within Preludium program (2025/57/N/ST2/03464) and from the Excellence Initiative - Research University of the Jagiellonian University in Krak\'ow. 
LLM artificial intelligence agents were used to assist the editing of this manuscript, but all scientific results, concepts, and claims represent original work by the human authors listed.

 \bibliography{References}

@article{romalis_hybrid_2010,
  title = {Hybrid Optical Pumping of Optically Dense Alkali-Metal Vapor without Quenching Gas},
  author = {Romalis, M. V.},
  journal = {Phys. Rev. Lett.},
  volume = {105},
  issue = {24},
  pages = {243001},
  numpages = {4},
  year = {2010},
  month = {Dec},
  publisher = {American Physical Society},
  doi = {10.1103/PhysRevLett.105.243001},
  url = {https://link.aps.org/doi/10.1103/PhysRevLett.105.243001}
}

@article{kopciuch_liouvillian_2025,
	title = {Liouvillian and {Hamiltonian} exceptional points of atomic vapors: {The} spectral signatures of quantum jumps},
	volume = {7},
	issn = {2643-1564},
	shorttitle = {Liouvillian and {Hamiltonian} exceptional points of atomic vapors},
	url = {https://link.aps.org/doi/10.1103/zxw5-nlsn},
	doi = {10.1103/zxw5-nlsn},
	language = {en},
	number = {3},
	urldate = {2026-04-14},
	journal = {Physical Review Research},
	author = {Kopciuch, Marek and Miranowicz, Adam},
	month = aug,
	year = {2025},
	pages = {033187},
}

@article{dembowski_encircling_2004,
	title = {Encircling an exceptional point},
	volume = {69},
	copyright = {http://link.aps.org/licenses/aps-default-license},
	issn = {1539-3755, 1550-2376},
	url = {https://link.aps.org/doi/10.1103/PhysRevE.69.056216},
	doi = {10.1103/PhysRevE.69.056216},
	
	number = {5},
	urldate = {2026-03-27},
	journal = {Physical Review E},
	author = {Dembowski, C. and Dietz, B. and Gräf, H.-D. and Harney, H. L. and Heine, A. and Heiss, W. D. and Richter, A.},
	month = may,
	year = {2004},
	pages = {056216},
}

@article{ellis_systematic_2025,
	title = {Systematic nonuniform field effects in miniature atomic co-magnetometers},
	volume = {10},
	issn = {2058-9565},
	url = {https://iopscience.iop.org/article/10.1088/2058-9565/adffb0},
	doi = {10.1088/2058-9565/adffb0},
	number = {4},
	urldate = {2026-03-27},
	journal = {Quantum Science and Technology},
	author = {Ellis, L M and Jayaseelan, M and Rushton, L M and Zipfel, J D and Bevington, P and Steele, B and Quick, G and Chalupczak, W and Guarrera, V},
	month = dec,
	year = {2025},
	pages = {045034},
}

@article{wiersig_review_2020,
	title = {Review of exceptional point-based sensors},
	volume = {8},
	issn = {2327-9125},
	url = {https://opg.optica.org/abstract.cfm?URI=prj-8-9-1457},
	doi = {10.1364/PRJ.396115},
	number = {9},
	urldate = {2026-03-27},
	journal = {Photonics Research},
	author = {Wiersig, Jan},
	month = sep,
	year = {2020},
	pages = {1457},
}

@article{bergholtz_exceptional_2021,
	title = {Exceptional topology of non-{Hermitian} systems},
	volume = {93},
	issn = {0034-6861, 1539-0756},
	url = {https://link.aps.org/doi/10.1103/RevModPhys.93.015005},
	doi = {10.1103/RevModPhys.93.015005},
	
	number = {1},
	urldate = {2026-03-27},
	journal = {Rev. Mod. Phys.},
	author = {Bergholtz, Emil J. and Budich, Jan Carl and Kunst, Flore K.},
	month = feb,
	year = {2021},
	pages = {015005},
}

@article{miri_exceptional_2019,
	title = {Exceptional points in optics and photonics},
	volume = {363},
	issn = {0036-8075, 1095-9203},
	url = {https://www.science.org/doi/10.1126/science.aar7709},
	doi = {10.1126/science.aar7709},
	
	number = {6422},
	urldate = {2026-03-27},
	journal = {Science},
	author = {Miri, Mohammad-Ali and Alù, Andrea},
	month = jan,
	year = {2019},
	pages = {eaar7709},
}

@article{parametrically_2018_Jiang,
    author = {Jiang, Liwei and Quan, Wei and Li, Rujie and Fan, Wenfeng and Liu, Feng and Qin, Jie and Wan, Shuangai and Fang, Jiancheng},
    title = {A parametrically modulated dual-axis atomic spin gyroscope},
    journal = {Applied Physics Letters},
    volume = {112},
    number = {5},
    pages = {054103},
    year = {2018},
    month = {01},
    issn = {0003-6951},
    doi = {10.1063/1.5018015},
    url = {https://doi.org/10.1063/1.5018015},
}

@article{Wang:2025lcm,
    author = "Wang, Hanfeng and Jacobs, Kurt and Fahey, Donald and Hu, Yong and Englund, Dirk R. and Trusheim, Matthew E.",
    journal={Nat. Phys.},
    volume={22},
    pages={577-583},
    title = "{Exceptional sensitivity near the bistable transition point of a hybrid quantum system}",
    doi={10.1038/s41567-026-03217-3},
    archivePrefix = "arXiv",
    primaryClass = "quant-ph",
    year = "2026"
}

@article{HEIL1995337,
title = {Very long nuclear relaxation times of spin polarized helium 3 in metal coated cells},
journal = {Phys. Lett. A},
volume = {201},
number = {4},
pages = {337-343},
year = {1995},
issn = {0375-9601},
doi = {https://doi.org/10.1016/0375-9601(95)00243-V},
url = {https://www.sciencedirect.com/science/article/pii/037596019500243V},
author = {Werner Heil and Hubert Humblot and Ernst Otten and Matthias Schafer and Reinhard Sarkau and Michèle Leduc},
}

@article{klingerPhysRevApplied.19.044092,
  title = {Optimization of Nuclear Polarization in an Alkali-Noble Gas Comagnetometer},
  author = {Klinger, Emmanuel and Liu, Tianhao and Padniuk, Mikhail and Engler, Martin and Kornack, Thomas and Pustelny, Szymon and Jackson Kimball, Derek F. and Budker, Dmitry and Wickenbrock, Arne},
  journal = {Phys. Rev. Appl.},
  volume = {19},
  issue = {4},
  pages = {044092},
  numpages = {11},
  year = {2023},
  month = {Apr},
  publisher = {American Physical Society},
  doi = {10.1103/PhysRevApplied.19.044092},
  url = {https://link.aps.org/doi/10.1103/PhysRevApplied.19.044092}
}

@article{Barbosa:2024vlt,
    author = "Barbosa, Alexandre and Ter{\c{c}}as, Hugo and Cruzeiro, Emmanuel Zambrini",
    title = "{Multiplexed quantum repeaters with hot multimode alkali-noble gas memories}",
    eprint = "2402.17752",
    archivePrefix = "arXiv",
    primaryClass = "quant-ph",
    month = "2",
    year = "2024",
    journal=""
}

@article{Allmendinger:2013eya,
    author = "Allmendinger, F. and Heil, W. and Karpuk, S. and Kilian, W. and Scharth, A. and Schmidt, U. and Schnabel, A. and Sobolev, Yu. and Tullney, K.",
    title = "{New Limit on Lorentz-Invariance- and CPT-Violating Neutron Spin Interactions Using a Free-Spin-Precession $^3$He - $^{129}$Xe Comagnetometer}",
    archivePrefix = "arXiv",
    primaryClass = "gr-qc",
    doi = "10.1103/PhysRevLett.112.110801",
    journal = "Phys. Rev. Lett.",
    volume = "112",
    number = "11",
    pages = "110801",
    year = "2014"
}

@article{GOLUB19941,
title = {Neutron electric-dipole moment, ultracold neutrons and polarized 3He},
journal = {Physics Reports},
volume = {237},
number = {1},
pages = {1-62},
year = {1994},
issn = {0370-1573},
doi = {https://doi.org/10.1016/0370-1573(94)90084-1},
url = {https://www.sciencedirect.com/science/article/pii/0370157394900841},
author = {R. Golub and Steve K. Lamoreaux}
}

@article{Tat:2025dyl,
    author = "Tat, Raymond",
    title = "{Demonstration of magic dressing of $^3$He}",
    eprint = "2512.02443",
    archivePrefix = "arXiv",
    primaryClass = "physics.atom-ph",
    month = "12",
    journal="",
    year = "2025"
}

@article{Walker1989PhysRevA,
  title = {Estimates of spin-exchange parameters for alkali-metal--noble-gas pairs},
  author = {Walker, Thad G.},
  journal = {Phys. Rev. A},
  volume = {40},
  issue = {9},
  pages = {4959--4964},
  numpages = {0},
  year = {1989},
  month = {Nov},
  publisher = {American Physical Society},
  doi = {10.1103/PhysRevA.40.4959},
  url = {https://link.aps.org/doi/10.1103/PhysRevA.40.4959}
}

@article{zhang2023search,
  title={Search for spin-dependent gravitational interactions at earth range},
  author={Zhang, S-B and Ba, Z-L and Ning, D-H and Zhai, N-F and Lu, Z-T and Sheng, Dong},
  journal={Phys. Rev. Lett.},
  volume={130},
  number={20},
  pages={201401},
  year={2023},
  publisher={APS}
}

@article{walker1997spin,
  title={Spin-exchange optical pumping of noble-gas nuclei},
  author={Walker, Thad G and Happer, William},
  journal={Rev. Mod. Phys.},
  volume={69},
  number={2},
  pages={629},
  year={1997},
  publisher={APS}
}

@inproceedings{walker2011fundamentals,
  title={Fundamentals of spin-exchange optical pumping},
  author={Walker, Thad G},
  booktitle={Journal of Physics: Conference Series},
  volume={294},
  number={1},
  pages={012001},
  year={2011}
}

@article{cong2025spin,
  title={Spin-dependent exotic interactions},
  author={Cong, Lei and Ji, Wei and Fadeev, Pavel and Ficek, Filip and Jiang, Min and Flambaum, Victor V and Guan, Haosen and Jackson Kimball, Derek F and Kozlov, Mikhail G and Stadnik, Yevgeny V and others},
  journal={Rev. Mod. Phys.},
  volume={97},
  number={2},
  pages={025005},
  year={2025},
  doi={10.1103/RevModPhys.97.025005},
  publisher={APS}
}

@article{katz2022optical,
  title={Optical quantum memory for noble-gas spins based on spin-exchange collisions},
  author={Katz, Or and Shaham, Roy and Reches, Eran and Gorshkov, Alexey V and Firstenberg, Ofer},
  journal={Phys. Rev. A},
  volume={105},
  number={4},
  pages={042606},
  year={2022},
  doi={10.1103/PhysRevA.105.042606},
  publisher={APS}
}

@article{katz2025optimal,
  title={Optimal strategies for optical quantum memories using long-lived noble-gas spins},
  author={Katz, Or and Reches, Eran and Shaham, Roy and Poem, Eilon and Gorshkov, Alexey V and Firstenberg, Ofer},
  journal={Phys. Rev. Research},
  volume={7},
  number={4},
  pages={043345},
  year={2025},
  doi={10.1103/xt9b-yq8q},
  publisher={APS}
}

@article{shaham2022strong,
  title={Strong coupling of alkali-metal spins to noble-gas spins with an hour-long coherence time},
  author={Shaham, Roy and Katz, Or and Firstenberg, Ofer},
  journal={Nature Phys.},
  volume={18},
  number={5},
  pages={506--510},
  year={2022},
  publisher={Nature Publishing Group UK London}
}

@article{su2022review,
  title={Review of noble-gas spin amplification via the spin-exchange collisions},
  author={Su, Haowen and Jiang, Min and Peng, Xinhua},
  journal={Sci. China Inf. Sci.},
  volume={65},
  number={10},
  pages={200501},
  year={2022},
  doi={10.1007/s11432-022-3550-1},
  publisher={Springer}
}

@article{allred2002high,
  title={High-sensitivity atomic magnetometer unaffected by spin-exchange relaxation},
  author={Allred, JCꎬ and Lyman, RN and Kornack, TW and Romalis, Michael V},
  journal={Phys. Rev. Lett.},
  volume={89},
  number={13},
  pages={130801},
  year={2002},
  publisher={APS},
  doi={10.1103/PhysRevLett.89.130801}
}

@book{happer2010optically,
  title={Optically pumped atoms},
  author={Happer, William and Jau, Yuan-Yu and Walker, Thad},
  year={2010},
  publisher={John Wiley \& Sons}
}

@article{haroche1970modified,
  title = {Modified Zeeman Hyperfine Spectra Observed in ${\mathrm{H}}^{1}$ and ${\mathrm{Rb}}^{87}$ Ground States Interacting with a Nonresonant rf Field},
  author = {Haroche, S. and Cohen-Tannoudji, C. and Audoin, C. and Schermann, J. P.},
  journal = {Phys. Rev. Lett.},
  volume = {24},
  issue = {16},
  pages = {861--864},
  numpages = {0},
  year = {1970},
  month = {Apr},
  publisher = {American Physical Society},
  doi = {10.1103/PhysRevLett.24.861},
  url = {https://link.aps.org/doi/10.1103/PhysRevLett.24.861}
}

@article{serafin2021squeezing,
  title = {Nuclear Spin Squeezing in Helium-3 by Continuous Quantum Nondemolition Measurement},
  author = {Serafin, Alan and Fadel, Matteo and Treutlein, Philipp and Sinatra, Alice},
  journal = {Phys. Rev. Lett.},
  volume = {127},
  issue = {1},
  pages = {013601},
  numpages = {6},
  year = {2021},
  month = {Jun},
  publisher = {American Physical Society},
  doi = {10.1103/PhysRevLett.127.013601},
  url = {https://link.aps.org/doi/10.1103/PhysRevLett.127.013601}
}

@article{put2019nonlinear,
  title = {Nonlinear magneto-optical rotation with parametric resonance},
  author = {Put, P. and Wcis\l{}o, P. and Gawlik, W. and Pustelny, S.},
  journal = {Phys. Rev. A},
  volume = {100},
  issue = {4},
  pages = {043838},
  numpages = {9},
  year = {2019},
  month = {Oct},
  publisher = {American Physical Society},
  doi = {10.1103/PhysRevA.100.043838},
  url = {https://link.aps.org/doi/10.1103/PhysRevA.100.043838}
}

@misc{wei_phase_2025,
	title = {Phase {Transition} {Control} in {Spin} {Ensemble} by {Spontaneous} {Symmetry} {Breaking}},
	copyright = {https://creativecommons.org/licenses/by/4.0/},
	url = {https://www.researchsquare.com/article/rs-7022021/v1},
	doi = {10.21203/rs.3.rs-7022021/v1},
	urldate = {2026-03-27},
	publisher = {In Review},
	author = {Wei, Kai and Wang, Weiyi and Lin, Shudong and Wang, Yuhao and Chai, Zhen and Fang, Jiancheng and Budker, Dmitry},
	month = nov,
	year = {2025},
}

@article{jiang_floquet_2021,
	title = {Floquet maser},
	volume = {7},
	copyright = {https://creativecommons.org/licenses/by-nc/4.0/},
	issn = {2375-2548},
	url = {https://www.science.org/doi/10.1126/sciadv.abe0719},
	doi = {10.1126/sciadv.abe0719},
	
	number = {8},
	urldate = {2026-01-09},
	journal = {Science Advances},
	author = {Jiang, Min and Su, Haowen and Wu, Ze and Peng, Xinhua and Budker, Dmitry},
	month = feb,
	year = {2021},
	pages = {eabe0719},
}

@article{jiang_floquet_2022,
	title = {Floquet {Spin} {Amplification}},
	volume = {128},
	issn = {0031-9007, 1079-7114},
	url = {https://link.aps.org/doi/10.1103/PhysRevLett.128.233201},
	doi = {10.1103/PhysRevLett.128.233201},
	
	number = {23},
	urldate = {2026-01-09},
	journal = {Physical Review Letters},
	author = {Jiang, Min and Qin, Yushu and Wang, Xin and Wang, Yuanhong and Su, Haowen and Peng, Xinhua and Budker, Dmitry},
	month = jun,
	year = {2022},
	pages = {233201},
}

@article{gavilan-martin_searching_2025,
	title = {Searching for dark matter with a spin-based interferometer},
	volume = {16},
	copyright = {2025 The Author(s)},
	issn = {2041-1723},
	url = {https://www.nature.com/articles/s41467-025-60178-6},
	doi = {10.1038/s41467-025-60178-6},
	
	number = {1},
	urldate = {2025-05-30},
	journal = {Nat. Commun.},
	publisher = {Nature Publishing Group},
	author = {Gavilan-Martin, Daniel and Łukasiewicz, Grzegorz and Padniuk, Mikhail and Klinger, Emmanuel and Smolis, Magdalena and Figueroa, Nataniel L. and Jackson Kimball, Derek F. and Sushkov, Alexander O. and Pustelny, Szymon and Budker, Dmitry and Wickenbrock, Arne},
	month = may,
	year = {2025},
	pages = {4953},
}

@article{jiang_search_2021,
	title = {Search for axion-like dark matter with spin-based amplifiers},
	volume = {17},
	copyright = {2021 The Author(s), under exclusive licence to Springer Nature Limited},
	issn = {1745-2481},
	url = {https://www.nature.com/articles/s41567-021-01392-z},
	doi = {10.1038/s41567-021-01392-z},
	
	number = {12},
	urldate = {2024-07-31},
	journal = {Nat. Phys.},
	publisher = {Nature Publishing Group},
	author = {Jiang, Min and Su, Haowen and Garcon, Antoine and Peng, Xinhua and Budker, Dmitry},
	month = dec,
	year = {2021},
	pages = {1402--1407},
}

@article{wei_ultrasensitive_2023,
	title = {Ultrasensitive {Atomic} {Comagnetometer} with {Enhanced} {Nuclear} {Spin} {Coherence}},
	volume = {130},
	url = {https://link.aps.org/doi/10.1103/PhysRevLett.130.063201},
	doi = {10.1103/PhysRevLett.130.063201},
	number = {6},
	urldate = {2024-07-31},
	journal = {Physical Review Letters},
	publisher = {American Physical Society},
	author = {Wei, Kai and Zhao, Tian and Fang, Xiujie and Xu, Zitong and Liu, Chang and Cao, Qian and Wickenbrock, Arne and Hu, Yanhui and Ji, Wei and Fang, Jiancheng and Budker, Dmitry},
	month = feb,
	year = {2023},
	pages = {063201},
}

@article{vasilakis_limits_2009,
    author = "Vasilakis, G. and Brown, J. M. and Kornack, T. W. and Romalis, M. V.",
    title = "{Limits on new long range nuclear spin-dependent forces set with a K - He-3 co-magnetometer}",
    archivePrefix = "arXiv",
    primaryClass = "physics.atom-ph",
    doi = "10.1103/PhysRevLett.103.261801",
    journal = "Phys. Rev. Lett.",
    volume = "103",
    pages = "261801",
    year = "2009"
}

@article{padniuk_universal_2024,
	title = {Universal determination of comagnetometer response to spin couplings},
	volume = {6},
	issn = {2643-1564},
	url = {https://link.aps.org/doi/10.1103/PhysRevResearch.6.013339},
	doi = {10.1103/PhysRevResearch.6.013339},
	
	number = {1},
	urldate = {2024-05-15},
	journal = {Physical Review Research},
	author = {Padniuk, Mikhail and Klinger, Emmanuel and Łukasiewicz, Grzegorz and Gavilan-Martin, Daniel and Liu, Tianhao and Pustelny, Szymon and Jackson Kimball, Derek F. and Budker, Dmitry and Wickenbrock, Arne},
	month = mar,
	year = {2024},
	pages = {013339},
}

@article{jackson2023probing,
    author = "Jackson Kimball, Derek F. and Budker, Dmitry and Chupp, Timothy E. and Geraci, Andrew A. and Kolkowitz, Shimon and Singh, Jaideep T. and Sushkov, Alexander O.",
    title = "{Probing fundamental physics with spin-based quantum sensors}",
    doi = "10.1103/PhysRevA.108.010101",
    journal = "Phys. Rev. A",
    volume = "108",
    number = "1",
    pages = "010101",
    year = "2023"
}

@article{bloch_axion-like_2020,
	title = {Axion-like relics: new constraints from old comagnetometer data},
	volume = {2020},
	issn = {1029-8479},
	shorttitle = {Axion-like relics},
	url = {https://doi.org/10.1007/JHEP01(2020)167},
	doi = {10.1007/JHEP01(2020)167},
	
	number = {1},
	urldate = {2022-11-03},
	journal = {Journal of High Energy Physics},
	author = {Bloch, Itay M. and Hochberg, Yonit and Kuflik, Eric and Volansky, Tomer},
	month = jan,
	year = {2020},
	pages = {167},
}

@misc{bloch_nasduck_2022,
	title = {{NASDUCK} {SERF}: {New} constraints on axion-like dark matter from a {SERF} comagnetometer},
	shorttitle = {{NASDUCK} {SERF}},
	url = {http://arxiv.org/abs/2209.13588},
	urldate = {2022-10-28},
	publisher = {arXiv},
	author = {Bloch, Itay M. and Shaham, Roy and Hochberg, Yonit and Kuflik, Eric and Volansky, Tomer and Katz, Or},
	month = sep,
	year = {2022},
	note = {arXiv:2209.13588 [hep-ex, physics:hep-ph, physics:physics, physics:quant-ph]},
}

@misc{lee_laboratory_2022,
	title = {Laboratory {Constraints} on the {Neutron}-{Spin} {Coupling} of {feV}-scale {Axions}},
	url = {http://arxiv.org/abs/2209.03289},
	urldate = {2022-10-27},
	publisher = {arXiv},
	author = {Lee, Junyi and Lisanti, Mariangela and Terrano, William A. and Romalis, Michael},
	month = sep,
	year = {2022},
	note = {arXiv:2209.03289 [astro-ph, physics:hep-ph, physics:physics]},
}

@article{safronova_search_2018,
	title = {Search for {New} {Physics} with {Atoms} and {Molecules}},
	volume = {90},
	doi = {10.1103/revmodphys.90.025008},
	number = {2},
	journal = {Rev. Mod. Phys.},
	author = {Safronova, Marianna and Budker, Dmitry and DeMille, David and Kimball, Derek F. Jackson and Derevianko, Andrei and Clark, Charles W.},
	month = jun,
	year = {2018},
	doi = {10.1103/revmodphys.90.025008},
	pages = {025008},
}

@article{padniuk_response_2022,
	title = {Response of atomic spin-based sensors to magnetic and nonmagnetic perturbations},
	volume = {12},
	issn = {2045-2322},
	url = {https://www.nature.com/articles/s41598-021-03609-w},
	doi = {10.1038/s41598-021-03609-w},
	
	number = {1},
	urldate = {2022-10-17},
	journal = {Sci. Rep.},
	author = {Padniuk, Mikhail and Kopciuch, Marek and Cipolletti, Riccardo and Wickenbrock, Arne and Budker, Dmitry and Pustelny, Szymon},
	month = dec,
	year = {2022},
	pages = {324},
}

@article{katz_coupling_2021,
	title = {Coupling light to a nuclear spin gas with a two-photon linewidth of five millihertz},
	volume = {7},
	url = {https://www.science.org/doi/10.1126/sciadv.abe9164},
	doi = {10.1126/sciadv.abe9164},
	number = {14},
	urldate = {2022-10-28},
	journal = {Science Advances},
	publisher = {American Association for the Advancement of Science},
	author = {Katz, Or and Shaham, Roy and Firstenberg, Ofer},
	month = apr,
	year = {2021},
	pages = {eabe9164},
}

@article{kornack_nuclear_2005,
	title = {Nuclear {Spin} {Gyroscope} {Based} on an {Atomic} {Comagnetometer}},
	volume = {95},
	url = {https://link.aps.org/doi/10.1103/PhysRevLett.95.230801},
	doi = {10.1103/PhysRevLett.95.230801},
	number = {23},
	urldate = {2022-10-17},
	journal = {Physical Review Letters},
	publisher = {American Physical Society},
	author = {Kornack, T. W. and Ghosh, R. K. and Romalis, M. V.},
	month = nov,
	year = {2005},
	pages = {230801},
}

@phdthesis{kornack_thomas_test_2005,
	type = {Phd {Thesis}},
	title = {A test of {CPT} and {Lorentz} symmetry using a potassium-helium-3 co-magnetometer},
	school = {Princeton University},
	author = {Kornack, Thomas},
	year = {2005},
}

@article{kornack_dynamics_2002,
	title = {Dynamics of {Two} {Overlapping} {Spin} {Ensembles} {Interacting} by {Spin} {Exchange}},
	volume = {89},
	issn = {0031-9007, 1079-7114},
	url = {https://link.aps.org/doi/10.1103/PhysRevLett.89.253002},
	doi = {10.1103/PhysRevLett.89.253002},
	
	number = {25},
	urldate = {2022-10-17},
	journal = {Physical Review Letters},
	author = {Kornack, T. W. and Romalis, M. V.},
	month = dec,
	year = {2002},
	pages = {253002},
}

@article{savukov_effects_2005,
	title = {Effects of spin-exchange collisions in a high-density alkali-metal vapor in low magnetic fields},
	volume = {71},
	url = {https://link.aps.org/doi/10.1103/PhysRevA.71.023405},
	doi = {10.1103/PhysRevA.71.023405},
	number = {2},
	urldate = {2022-10-17},
	journal = {Physical Review A},
	publisher = {American Physical Society},
	author = {Savukov, I. M. and Romalis, M. V.},
	month = feb,
	year = {2005},
	pages = {023405},
}

@article{shaham_strong_2022,
	title = {Strong coupling of alkali-metal spins to noble-gas spins with an hour-long coherence time},
	volume = {18},
	issn = {1745-2473, 1745-2481},
	url = {https://www.nature.com/articles/s41567-022-01535-w},
	doi = {10.1038/s41567-022-01535-w},
	
	number = {5},
	urldate = {2022-10-17},
	journal = {Nat. Phys.},
	author = {Shaham, R. and Katz, O. and Firstenberg, O.},
	month = may,
	year = {2022},
    doi={10.1038/s41567-022-01535-w},
	pages = {506--510},
}

@article{li_parametric_2006,
	title = {Parametric modulation of an atomic magnetometer},
	volume = {89},
	issn = {0003-6951, 1077-3118},
	url = {http://aip.scitation.org/doi/10.1063/1.2357553},
	doi = {10.1063/1.2357553},
	
	number = {13},
	urldate = {2023-01-23},
	journal = {Applied Physics Letters},
	author = {Li, Zhimin and Wakai, Ronald T. and Walker, Thad G.},
	month = sep,
	year = {2006},
	pages = {134105},
}

@article{happer_spin-exchange_1973,
	title = {Spin-{Exchange} {Shift} and {Narrowing} of {Magnetic} {Resonance} {Lines} in {Optically} {Pumped} {Alkali} {Vapors}},
	volume = {31},
	issn = {0031-9007},
	url = {https://link.aps.org/doi/10.1103/PhysRevLett.31.273},
	doi = {10.1103/PhysRevLett.31.273},
	
	number = {5},
	urldate = {2022-12-12},
	journal = {Physical Review Letters},
	author = {Happer, W. and Tang, H.},
	month = jul,
	year = {1973},
	pages = {273--276},
}

@article{happer_effect_1977,
	title = {Effect of rapid spin exchange on the magnetic-resonance spectrum of alkali vapors},
	volume = {16},
	issn = {0556-2791},
	url = {https://link.aps.org/doi/10.1103/PhysRevA.16.1877},
	doi = {10.1103/PhysRevA.16.1877},
	
	number = {5},
	urldate = {2022-12-05},
	journal = {Physical Review A},
	author = {Happer, W. and Tam, A. C.},
	month = nov,
	year = {1977},
	pages = {1877--1891},
}

@article{appelt_theory_1998,
	title = {Theory of spin-exchange optical pumping of 3 {He} and 129 {Xe}},
	volume = {58},
	issn = {1050-2947, 1094-1622},
	url = {https://link.aps.org/doi/10.1103/PhysRevA.58.1412},
	doi = {10.1103/PhysRevA.58.1412},
	
	number = {2},
	urldate = {2022-12-01},
	journal = {Physical Review A},
	author = {Appelt, S. and Baranga, A. Ben-Amar and Erickson, C. J. and Romalis, M. V. and Young, A. R. and Happer, W.},
	month = aug,
	year = {1998},
	pages = {1412--1439},
}

@article{katz_quantum_2022,
	title = {Quantum {Interface} for {Noble}-{Gas} {Spins} {Based} on {Spin}-{Exchange} {Collisions}},
	volume = {3},
	issn = {2691-3399},
	url = {https://link.aps.org/doi/10.1103/PRXQuantum.3.010305},
	doi = {10.1103/PRXQuantum.3.010305},
	
	number = {1},
	urldate = {2022-10-17},
	journal = {PRX Quantum},
	author = {Katz, Or and Shaham, Roy and Firstenberg, Ofer},
	month = jan,
	year = {2022},
	pages = {010305},
}

@article{Zhao:2025wvv,
    author = "Zhao, Yixin and Peng, Xiang and Wu, Teng and Guo, Hong",
    title = "{Response of comagnetometers to exotic spin-dependent couplings}",
    archivePrefix = "arXiv",
    primaryClass = "hep-ph",
    doi = "10.1103/824f-v3w5",
    journal = "Phys. Rev. D",
    volume = "112",
    number = "1",
    pages = "012023",
    year = "2025"
}
 \newpage
 
\appendix
\section{Derivation of the Bessel renormalization of the transverse coupling}
\label{app:matrix_derivation}

Here we show how the interaction matrix introduced in Eq.~(\ref{eq:interaction_matrix}) is transformed into the effective matrix of Eq.~(\ref{eq:effective_matrix}) under longitudinal field modulation.

Starting from Eq.~(\ref{eq:interaction_matrix}), we remove the explicit modulation from the diagonal terms by moving to a rotating frame defined by the unitary transformation

\begin{equation}
U(t) = 
\begin{pmatrix}
e^{-i \phi_e(t)} & 0 \\
0 & e^{-i \phi_n(t)}
\end{pmatrix}
\end{equation}
with phases
\begin{align}
\phi_e(t) &=   \frac{\Omega_e}{\omega_{\mathrm{m}}} \sin(\omega_{\mathrm{m}} t)\,, \\
\phi_n(t) &=   \frac{\Omega_n}{\omega_{\mathrm{m}}} \sin(\omega_{\mathrm{m}} t)\,.
\end{align}

The transformed interaction matrix is 
\begin{equation}
\tilde{M}(t) = U^{\dagger} M U - i\, U^{\dagger} \frac{d}{dt} U\,.
\end{equation}
With this choice of phases, the oscillating diagonal terms are cancelled, but the modulation is absorbed into off-diagonal elements. The interaction matrix now reads:
\begin{equation}
\tilde{M}(t) =
\begin{pmatrix}
\omega_e-i\Gamma_e &
\eta_{en}\, e^{-i[\phi_e(t)-\phi_n(t)]} \\
\eta_{ne}\, e^{+i[\phi_e(t)-\phi_n(t)]} &
\omega_n-i\Gamma_n
\end{pmatrix}.
\end{equation}


Using
\begin{equation}
\phi_e(t)-\phi_n(t)
=
\frac{\Omega_e-\Omega_n}{\omega_m}\sin(\omega_m t)
\equiv \beta \sin(\omega_m t),
\end{equation}
we rewrite the matrix as
\begin{equation}
\label{eq:matrix_before_Jacobi_expansion}
\tilde{M}(t) =
\begin{pmatrix}
\omega_e-i\Gamma_e &
\eta_{en}\, e^{-i\beta \sin(\omega_m t)} \\
\eta_{ne}\, e^{+i\beta \sin(\omega_m t)} &
\omega_n-i\Gamma_n
\end{pmatrix},
\end{equation}

Now we can use the Jacobi-Anger expansion 
\begin{equation}
\label{eq:jacobi_expansion}
e^{i \beta \sin(\omega_{\mathrm{z}} t)} = \sum_{n=-\infty}^{\infty} J_n(\beta) \, e^{i n \omega_{\mathrm{z}} t}\,.+
\end{equation}
Substituting this expansion into the off-diagonal terms shows that the coupling is decomposed into harmonics oscillating at integer multiples of the modulation frequency.

In the fast modulation regime, where $\omega_m$ is much faster than the intrinsic dynamical rates of the coupled-spin system ($\omega_e$, $\omega_n$, $\eta=\sqrt{\eta_{en}\,\eta_{ne}}$, $\Gamma_e$, $\Gamma_n$)
, the rapidly oscillating terms with $k\neq 0$ average to zero over one modulation cycle. Retaining only the secular contribution, corresponding to $k=0$, yields the effective time-averaged interaction matrix
\begin{equation}
M_{\mathrm{eff}}(\beta)=
\begin{pmatrix}
\omega_e-i\Gamma_e &
\eta_{en}\,J_0(\beta)\\[6pt]
\eta_{ne}\,J_0(\beta) &
\omega_n-i\Gamma_n
\end{pmatrix}.
\end{equation}

The longitudinal parametric modulation renormalizes the transverse exchange coupling by the Bessel function $J_0(\beta)$, while leaving the longitudinal mean-field shifts unchanged.

\section{Effects of parametric modulation on alkali-metal spins} \label{app:alkali metal modAmp}

The alkali-metal ensemble provides an indirect readout of the coupled spin system through Faraday rotation of a linearly polarized probe beam propagating along the $x$-direction and can be treated, to a good approximation, as a standard SERF magnetometer. The effect of parametric modulation on the transverse response of optically pumped alkali spins was studied in Ref.~\cite{li_parametric_2006}. Here we directly adopt their results for the experimentally relevant slow components of the signal at the first and second harmonics of the modulation frequency, which govern the comagnetometer readout:

\begin{equation}
P_x^e(\omega_m) \approx \frac{\frac{\gamma_e}{q}\alpha_e b_\perp^x \, P^e_{\parallel}}{\Gamma_e}\, 2 J_0(\beta)J_1(\beta)\,\sin(\omega_m t)\,,
\label{eq:demodsignalx}
\end{equation}

\begin{equation}
P_y^e(2\omega_m) \approx \frac{\frac{\gamma_e}{q}\alpha_e b_\perp^y \, P^e_{\parallel}}{\Gamma_e} \, 2 J_0(\beta)J_2(\beta)\, \cos(2\omega_m t)\,.
\label{eq:amplitude first harmonic}
\end{equation}

Notably, the decoupled regime coincides with vanishing alkali response when $J_0(\beta)\rightarrow 0$.

 \section{Experimental setup}\label{app:ExpSetup}

\begin{figure}[htbp]
    \centering
    \includegraphics[width=0.75\linewidth]{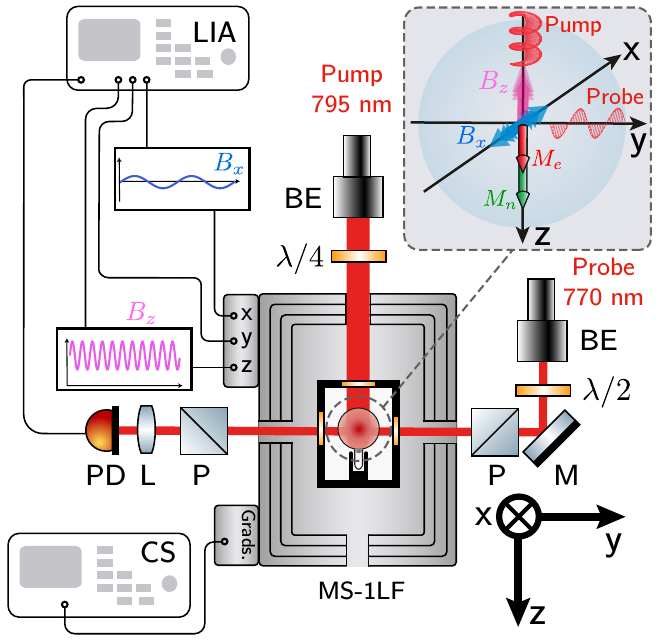}
    \caption{Sketch of the experimental setup. A vapor cell containing $^3$He, N$_2$, $^{87}$Rb, and K is situated inside a Twînleaf MS-1LF magnetic shield. The Rb atoms are optically pumped with circularly-polarized light in resonance with the Rb D1 line. Potassium (and then $^3$He) spins are pumped by spin-exchange collisions with Rb. The readout is realized by measuring the polarization rotation of a linearly polarized probe beam detuned from the K D1 line. Parametric modulation is applied by modulating the leading field with a sine wave and the resulting signal demodulated using a lock-in amplifier (LIA). Low-frequency modulation at 40\;Hz of $B_x$ allows for closed-loop control of the compensation point. The gradient coils are controlled with a low-noise current source (CS).}
    \label{fig:setup}
\end{figure}

The schematic of the comagnetometer setup is shown in Fig.\;\ref{fig:setup}. At the core of the comagnetometer is a 20-mm spherical vapor cell (Twînleaf) filled with 3 amg of $^3$He as the nuclear spin content. To prevent radiation trapping, the cell also contains about 50\;Torr of N$_2$. As for the electronic spin content, the cell is loaded with a mixture of $^{87}$Rb (1\%) and natural abundance K (99\%) atoms, allowing for reduced light-shift sensitivity through hybrid optical pumping \cite{romalis_hybrid_2010}. To reach the operating temperature of 185\;$^\circ$C, the cell is placed in an oven and heated with AC resistive heaters. The oven assembly is mounted in a Twînleaf MS-1LF 4-layer magnetic shield. Built-in flexible printed circuits allow, on the one hand, to apply necessary magnetic fields in all three directions and on the other hand to compensate for potential magnetic field gradients.  The spatial gradients inside the vapor cell are compensated following the method  Ref.\,\cite{klingerPhysRevApplied.19.044092}, allowing for longer nuclear spins coherence times and thus improving the stability of system.
The comagnetometer readout is realized by measuring the polarization rotation of a linearly-polarized probe beam detuned about 0.5\,nm towards the shorter wavelength of the K D1 line.

The comagnetometer is operated in the self-compensating regime \cite{kornack_nuclear_2005,klingerPhysRevApplied.19.044092} by setting the leading field $B_z$ to 
\begin{equation}
B_{c}=-(B_\parallel^e + B_\parallel^n) 
\,.
\label{eq:compensation point}
\end{equation}
At this specific magnetic field, $^3$He spins adiabatically compensate for low-frequency (below $^3$He precession frequency) magnetic field changes orthogonal to the leading field.

Another point of interest is the hybrid-resonance point $B_r$ \cite{shaham_strong_2022}, where the bare precession frequencies of both nuclear and electron spins are equal ($\omega_e=\omega_n$). This is attained by setting the leading field to
\begin{equation}
    B_r = \frac{\gamma_n B_\parallel^e - \gamma_e B_\parallel^n}{\gamma_e - \gamma_n}
    \,.
\end{equation}
    
In the case of a K-$^3$He comagnetometer, the effective magnetic field of the electron spins is a small fraction of the nuclear effective field: $|B^n| \gg |B^e| $. Because of this, the hybrid resonance point and the self-compensation point are almost degenerate, $B_c\approx B_r\approx  -B^n_\parallel $. Operating the comagnetometer near these points is beneficial for precision measurements (see, e.g., Refs.\,\cite{vasilakis_limits_2009,lee_laboratory_2022,gavilan-martin_searching_2025}) as adiabatic changes in the magnetic field are compensated, and transverse $^3$He precession is dampened compared to a free nuclear spin system with no coupling ($\eta=0$).

Parametric modulation is applied by modulating the leading field with a sine wave, and the resulting signal is demodulated using a lock-in amplifier. The demodulated signal has two quadratures:
$A_x=A\cos(\phi_m)$ and $A_y=A\cos(\phi_m)$, with $\phi_m$ the demodulation phase.
We are not interested in $\phi_m$, and it changes when $B_m$ varies. For the purpose of fitting the FPD with \cref{eq:timeseries fit} in \cref{sec:results}, we fit $R=\sqrt{(A_X+a)^2+(A_Y+a)^2}$, removing the phase $\phi_m$. The parameter $a$ is an offset large enough to avoid zero crossings of the signal that would distort the FPD.
\newline

\section{Parametric modulation effects on self-compensating comagnetometer} \label{app:self-compensatig response}

Hot alkali-metal vapors containing mixtures of alkali-metal and noble-gas atoms are commonly used as self-compensating comagnetometers, exploiting the unique properties of the so-called compensation point. The compensation point is defined as a specific value of the longitudinal magnetic field at which the applied field offsets the sum of the Fermi-contact fields generated by the two spin ensembles, see \cref{eq:compensation point}. At this operating point, the response of the system to transverse magnetic fields is strongly suppressed, while sensitivity to non-magnetic couplings, including rotation  or exotic spin interactions , is preserved \cite{kornack_nuclear_2005}. 

The calibration of comagnetometers to non-magnetic interacions have been addressed in several studies
\cite{padniuk_response_2022,kornack_thomas_test_2005,padniuk_universal_2024,Zhao:2025wvv}
particularly in the context of exotic spin couplings, for which the response cannot be directly measured.
While the present work does not aim to investigate self-compensating comagnetometer operation experimentally, and comagnetometric applications lie outside its primary scope, we believe it is nevertheless useful to make explicit how parametric modulation modifies the frequency-response function in the vicinity of the compensation point.
We build on the previous work of Ref.~\cite{padniuk_response_2022} and extend the frequency-response formalism to account for modulation-induced decoupling of the transverse spin components. The frequency-dependent response function of the alkali polarization is given by: 
\begin{widetext}
    \begin{equation}
         P_\perp^{e\pm}(\omega) = -\frac{\gamma_eP^e_\parallel b_\perp(\omega)}{q} \frac{\omega_n^{\prime}(\alpha_e-J_0(\beta)\alpha_n)+(\pm\omega+\gamma_n\Delta_{B_z}-i \Gamma_n)\alpha_e}{(\pm\omega+\omega^{\prime}_e+\gamma_e\Delta_{B_z}/q-i\Gamma_e/q)(\pm \omega +\omega^{\prime}_n-+\gamma_n \Delta_{B_z}-i\Gamma_n)-J_0^2(\beta)\omega^{\prime}_e\omega^{\prime}_n}\,,
    \label{eq:comag_frequency_response}
    \end{equation}
\end{widetext}
where $\omega^{\prime}_e = \gamma_e\lambda M^e P^e_\parallel/q$ and $\omega^{\prime}_n = \gamma_n\lambda M^nP^n_\parallel$ are the Larmor frequencies of the electron and nuclear polarizations at the compensation point, and $b_\perp(\omega)$ is the spectral amplitude of the perturbation. 

\section{Broadening at large $B_m$ } \label{app:high Bz}
Large $B_m$ can lead to a reduction of the amplitude of the response of the system and an increase in the electronic relaxation rate $\Gamma_e$. When the leading field modulation amplitude $B_\text{m}$ is large compared to the static bias field $B_z$, there is additional relaxation of the electronic polarization, resulting in a reduced response amplitude. Even if, on average, the applied field due to parametric modulation is zero, some effects depend on the square of the applied magnetic field, and does not average out. This is the case of the resonance broadening due to a large field applied in the leading direction, which can be written as \cite{klingerPhysRevApplied.19.044092,appelt_theory_1998}
\begin{equation}
    \Gamma_{m}=\frac{5}{36} \frac{\left(\gamma_e B_m\right)^2}{q^2 R_{\mathrm{se}}},
\end{equation}
where $R_{\mathrm{se}}$ is the alkali-alkali spin-exchange rate.

To explore high values of the modulation index $\beta$ large  $B_m$ is applied. This leads to a high $\Gamma_m$ such that the leading relaxation mechanism is $\Gamma_e \approx \Gamma_m$. Then, not only the Bessel functions $J_0(\beta),J_1(\beta)$ are a function of the modulation amplitude $B_m$, but the electronic relaxation rate itself.
In particular, the explicit dependence on $\Gamma_m$ for the amplitude can be obtained using \cref{eq:demodsignalx}, $P^e_x\propto 1/B_m^2$. The nuclear decay rate dependence on $B_m$ is obtained via \cref{eq:lambda effective}, $\Gamma_n \propto \frac{\Gamma_m}{(\Delta\omega)^2+\Gamma_m^2}$.
\begin{figure}[htb]
    \centering
    \includegraphics[width=1\linewidth]{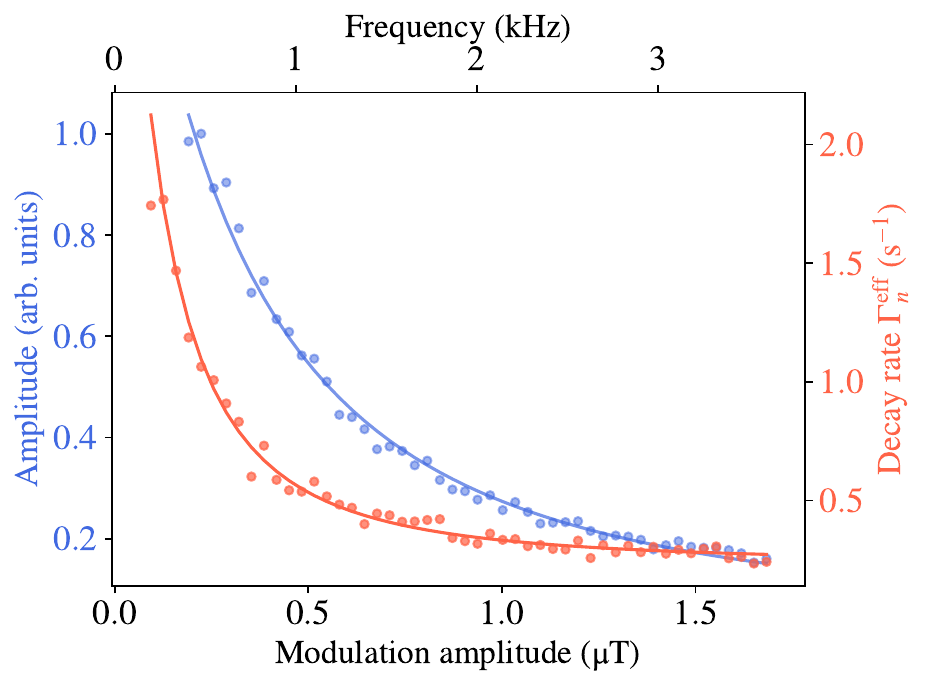}
    \caption{Amplitude and nuclear decay rate for fixed modulation index $\beta$. By increasing the amplitude and frequency of the modulation proportionally, we can characterize the suppression of the response and the decay due to high applied magnetic fields. Experimentally measured data points are fitted with \cref{eq:demodsignalx,eq:lambda effective}, respectively, assuming $\Gamma_e=\Gamma_m$. The fits are used as input when fitting \cref{eq:beta fits}.
    }
    \label{fig:fixed ratio}
\end{figure}

To account for this effect, we measure the FPD response of the system maintaining a fixed ratio of $B_m/\omega_m$, which fixes the argument of the Bessel functions $\beta$ over the increasing modulation amplitude,  according to the model (\cref{sec:theory}). If $\Gamma_e\gg\Gamma_m$ there would be a flat dependence on both amplitude and decay rate for the fixed ratio, \cref{eq:omega effective}.  However, this is not the case, as it is shown in the results of the fits (see \cref{fig:fixed ratio}). Both the amplitude and the decay rate can be fit by \cref{eq:demodsignalx,eq:lambda effective}, respectively, assuming $\Gamma_e=\Gamma_m$.
We obtain expressions $\Gamma(B_m)$ and $A(B_m)$ that take into account the effects that do not depend on the modulation index $\beta$ and are used when fitting \cref{eq:beta fits}.

\section{Parametric modulation effects away from hybrid resonance}
\label{app:off-resonance-coupling}

The derivation presented in \cref{app:matrix_derivation} can be extended beyond the slow dynamics ($k=0$) approximation. Starting from the rotating-frame interaction matrix in Eq. (\ref{eq:matrix_before_Jacobi_expansion}) and using the Jacobi--Anger expansion in Eq. (\ref{eq:jacobi_expansion})
the off-diagonal coupling is decomposed into a set of harmonics oscillating at integer multiples of the modulation frequency.

When the bare detuning $\Delta\omega=\omega_e-\omega_n$ is large compared to the coupling strength $\eta$, the $k=0$ term describes a weak, off-resonant interaction. However, if the modulation frequency satisfies the resonance condition
\begin{equation}
\Delta\omega \approx k\omega_m,
\end{equation}
the $k$-th harmonic becomes slowly varying and can mediate resonant coupling between the two spin ensembles.

To isolate this contribution, we move to a frame rotating at $\pm k\omega_m/2$, yielding the effective interaction matrix
\begin{equation}
M_{\mathrm{eff}}^{(k)}=
\begin{pmatrix}
\omega_e-\frac{k\omega_m}{2}-i\Gamma_e &
\eta_{en}\,J_k(\beta)\\[6pt]
\eta_{ne}\,J_k(\beta) &
\omega_n+\frac{k\omega_m}{2}-i\Gamma_n
\end{pmatrix}.
\end{equation}
The corresponding eigenvalues take the form
\begin{equation}
\begin{split}
\xi_\pm^{(k)} &=
\frac{\omega_e+\omega_n}{2}
-i\frac{\Gamma_e+\Gamma_n}{2}
\\
&\quad\pm
\sqrt{
\frac{(\Delta \omega - k\omega_\text{m}-i\Delta\Gamma)^2}{4}
+\eta^2 J_k^2(\beta)
},
\end{split}
\end{equation}
where $\eta^2=\eta_{en}\eta_{ne}$ and $\Delta\Gamma=\Gamma_e-\Gamma_n$.

This result shows that longitudinal modulation can restore strong effective coupling even far from the bare hybrid resonance. In this regime, the interaction is mediated by Floquet sidebands and has an effective strength $\eta_{\mathrm{eff}}^{(k)}=\eta J_k(\beta)$. 

 \end{document}